\documentclass[11pt]{article}
\usepackage{amsmath}
\usepackage{amssymb}
\usepackage{authblk}

\usepackage{ifpdf}
\ifpdf 
    \usepackage[pdftex]{graphicx}   
    \usepackage[pdftex,     
            plainpages=false,   
            breaklinks=true,    
            colorlinks=true,
            pdftitle=My Document
            pdfauthor=My Good Self
           ]{hyperref} 
    \usepackage{thumbpdf}
\else 
    \usepackage{graphicx}       
    \usepackage{hyperref}       
\fi 

\title{A Survey of Selected Algorithms Used in Military Applications
from the Viewpoints of Dataflow and GaAs
}
\author[1]{Ilir Capuni}
\author[2]{Veljko Milutinovi\'{c}}
\affil[1]{Department of Computer Science, Barleti University, Tirana, Albania }
\affil[2]{Department of Computer Science, University of Indiana, Bloomington, IN, USA}

\date{}

\begin{document}
\maketitle

\begin{abstract}
This is a short survey of ten algorithms that are often used for military purposes, 
followed by analysis of their potential suitability for dataflow and GaAs, 
which are a specific architecture and technology for supercomputers on a chip, respectively.

Whenever an algorithm or a device is used in military settings, it is natural to assume strict 
requirements related to speed, reliability, scale, energy, size, and accuracy. 
The two aforementioned paradigms seem to be promising in fulfilling most of these requirements.

 \end{abstract}

\section{Introduction}

This is a mini survey of ten specific algorithms for optimization and learning, 
combined with an analysis of their suitability for dataflow implementations in 
future supercomputers on a chip, as described 
in~\cite{vmUltimateDataFlow}, and their suitability for GaAs technology, which is
 also an option for future supercomputers on a chip.

In section 2, each algorithm is specified with appropriate mathematical 
and logical notions, and presented using the guidelines from
~\cite{vm1996best},  which means the
following:
\begin{enumerate}
\item  What is the problem to be solved.
\item  What was the best existing algorithm
     prior to the introduction of the presented one.
\item  Why the newly proposed algorithm is better?
\item  How much is it better and under which conditions?
\end{enumerate}

Each of the above issues is presented as concisely as possible to 
provide an effective insight into the essence.

In section 3, each algorithm is described from the viewpoint of its suitability 
for dataflow technology, which is of interest for high-speed, high- precision, 
low-power, and low-size applications in aerospace and defense.

In section 4, each algorithm is described from the view-point of its suitability 
for GaAs technology, which is of interest for ultra-high-speed processing in 
high radiation environments, typical of aerospace and defense.

Section 5 presents conclusions related to the price/performance ratio for 
various scenarios of interest for this survey.

The last part includes references. Four of them are general,   while the 
others are related to the ten surveyed algorithms.

\section{Algorithms}
Artificial intelligence (AI) and algorithms in general have made their way into 
the area of defense, where their widespread use is changing the classical 
doctrines of warfare and defense in general. 
Algorithms are used in detection, planning, field operations, and support 
functions, which are the main tasks in the defense sector. 
Algorithms and smart sensors are used to detect potentially dangerous
 persons and objects at border crossings, customs checkpoints, and other ports of travel. 
 The insights gained as outputs of these algorithms are used to deploy active policing 
 and provide a more holistic understanding of crisis scenarios. In planning, available 
 data and algorithms are used to better anticipate resourcing requirements and 
 associated costs for missions and training exercises. 
 In field operations, these can provide real-time information and quick assessment 
 to improve mission outcomes, protect people, assets, and information.
Some systems and weapons are equipped with various support and decision-making
 systems while unmanned vehicles and robots perform tasks that involve safety risks 
 with high accuracy and fewer resources.

In this paper we present ten algorithms that are used in military applications for core,
 tactical, and support operations. Instances of underlying problems, i.e., inputs for 
 algorithms, are usually large in scale. 
 Therefore, there is a large amount of data that need to be processed fast, preferably
  in real time, accurately and reliably, while guaranteeing confidentiality and control of 
  information even when operating in adverse conditions.

The most frequently found algorithms in DARPA sponsored projects are related to the 
problems and algorithms that we are presenting below. The selection of algorithms and 
problems seems to be slightly biased towards computer vision for it seems to be the
 essential technology for the development of autonomous vehicles, replacing the driver’s 
 eye, thus allowing the   vehicle to detect objects of interest in dangerous locations.

\subsection{Large-scale stochastic programming problems}

There are plenty of optimization problems occurring on a daily basis in the military with some uncertainties 
that are usually represented as scenarios. Applications range from scheduling monthly or 
daily air or sea lifts with uncertain 
cargo~\cite{attackedsealift}, to cyber workforce planning~\cite{ArmyCyberWorkforcePlanner} 
or planning of medical facility deployment~\cite{medicalallocation}.  
These need fast, accurate, and reliable solvers of large-scale stochastic programming problems. 

\emph{Benders decomposition} is a mathematical programming technique for solving very large 
linear programming problems with specific block structure~\cite{benders1962partitioning}.
\\ Suppose that a problem occurs in two or more stages, where the decisions for the later stages 
depend on the results from the previous ones. 
The first attempt to make a decision for the first stage problem is performed without
 prior knowledge of optimality regarding later stage decisions. The first stage decision 
 is the master problem, whereas subsequent stages are considered as separate subproblems
  whose information is passed back to the master problem.
   If any violation of a constraint of a subproblem is detected, the constraint is
    added back to the master problem, which is then resolved. 
    The master problem represents an initial convex set which is further constrained 
    by information gathered from the subproblems hence shrinking the feasible 
    space as information is added.
If matrices $A$ and $B$ represent the constraints, and $Y$ represents the feasible set of $\vec{y}$, 
the problem we are solving is represented as 
a minimization problem as follows

\begin{align*}
& \text{minimize} && \vec{c}^\top\vec{x} + \vec{d}^\top\vec{y} \\
& \text{subject to} && A \vec{x} + B \vec{y} \geq \vec{b} \\
& && y \in Y \\
& && \vec{x} \geq \vec{0}
\end{align*} 

Linear programming in general is an NP-complete problem and Benders decomposition
 is a method that slowly converges to the desired solution. It decomposes the problem into a large
 number of smaller problems that are essentially polynomial-time computable while 
 their independence allows for efficient use of parallelism.

\subsection{Image registration }

The absolute accuracy of robotic arms and autonomous armored vehicles is 
possible thanks to computer vision applied on high-resolution inputs gathered from 
multiple cameras which ultimately need to be transformed into the same coordinate 
system to create consistent data for subsequent algorithms. 
This is done using the so-called ~\emph{image registration algorithm}. 

Image registration involves spatially transforming the 
source image(s) to align with the target image.
The alignment is performed using a specific mapping called 
\emph{homography} which is defined as follows. 

Let $(x,y)$ be a 2D point in an image. It can be represented as a 3D 
vector $\vec{x}=(x_1, x_2, x_3)$, where $x=x_1/x_3$ and $y = x_2/x_3$ which is 
actually a point on a projective plane. 
Let $f(x,y)$ and $f'(x,y)$ be two images. We need to estimate a transformation $T$ such that 
$$f(x,y) = f'(T(x,y)).$$ 

There is a plethora of algorithms that accurately perform image registration. 
When images are exposed to noise causing a scene to appear cluttered in an image, 
the best existing solution has polynomial time complexity (see ~\cite{polyreg}). 

\subsection{Video stitching}

Image and video stitching is the process of removing the limitations of the field of 
vision in an image or video by stitching several multiple overlapping images/videos to 
obtain a wide field of image/video view. 
Video stitching
is essentially a generalization of a multi-image stitching with a new set of constraints and challenges.

First, images need to be transformed into the same coordinate system using previously 
explained image registration. 
Then, depending on the use case, an appropriate algorithm is chosen to find the 
seams of the stitches.

Faster and earlier developed algorithms usually perform global deformation 
and alignment of multiple overlapping images according to an estimated single transformation.
Recently developed algorithms transform the problem into more advanced 
optimization problems that consider camera movement, which requires 
stabilization of the video. 
Video stitching can therefore be posed as 
optimizing an objective function consisting of a stabilization term and a stitching 
term on which an iterative optimization algorithm is performed. Clearly, vast amount of 
computation is required even for stitching low-resolution videos
 (see~\cite{ImageVideoStitching}).

\subsection{Pattern recognition algorithms}

Pattern recognition is concerned with automatic identification of 
regularities in data and classifying data into different categories.

Formally,  given an unknown function 
$g:{\mathcal {X}}\rightarrow {\mathcal {Y}}$
 (the ground truth) that maps input instances  $\boldsymbol {x}\in {\mathcal {X}}$ 
 to output labels  $y \in \mathcal{Y}$, 
 along with training data 
 $\mathbf {D} =\{({\boldsymbol {x}}_{1},y_{1}),\dots ,({\boldsymbol {x}}_{n},y_{n})\}$ 
 assumed to represent accurate examples of the mapping, 
 produce a function $h:{\mathcal {X}}\rightarrow {\mathcal {Y}}$ 
 that approximates the correct mapping $g$ as closely as possible. 

\emph{Classification} is the problem of identifying to which of a set of categories 
(sub-populations) an observation (or observations) belongs. 
Examples for this are labeling an identified object on a video stream as an enemy, 
or assigning a device in a network a diagnosis based on observed characteristics 
of the device (type, presence of certain features, etc.). For the classification problem 
we will consider logistic regression, kNN, perceptrons, and SVM. 

\emph{Clustering} is a method for classifying and predicting categorical labels, 
and for this category we will present the k-means algorithm.

Finally, we will also consider the \emph{ensemble learning}, whose essence are 
supervised meta-algorithms for combining multiple learning algorithms together.

\subsubsection{Logistic Regression}

The logistic model is used in statistics to assess the probability of a certain 
event or an existing set of them such as pass/fail, win/lose, alive/dead, 
or enemy/friend. It can also be extended to model several classes of events 
such as determining if a person in an image has an RPG or not, 
if an image contains a specific object, etc. Each object detected in an image 
is assigned probability between 0 and 1, with the sum of one.

 Formally, consider a single training data point $(\mathbf{x}, y )$
 and set 
 $$P(Y=1| \mathbf{X} = x) = \theta_0 + \sum_{i=1}^m \theta_i x_i.$$

Now, using the maximum likelihood estimator or even the Newton method, 
find the values of $\theta_i$, $i=0, 1, \dots, m$, that maximize  
the probability for all the data. 

Calling 
$n$
 the size of a training sample and $m$ the number of weights, 
 training will take $O(m^2n + m^3)$ steps and prediction $O(m)$ steps.

\subsubsection{Online kNN}

kNN is a supervised learning algorithm that stores the labeled trained 
examples given as  pairs 
$(X_1,Y_1), (X_2,Y_2), \dots, (X_n, Y_n)$ 
taking values in $\mathbb{R}^d \times \{1,2\}$.

Training stage consists of just storing these samples. 
To make a prediction, kNN algorithms find
the $k$ nearest neighbors of a query point and compute 
the class label based on the $k$ nearest most similar points. 

To compute distance to one example $O(d)$ steps are needed. 
 $O(nd)$ steps are needed to find the nearest neighbor. 
$O(nkd)$ steps are needed to find $k$ closest examples. 
This is is prohibitively expensive for a large number of samples. 

Faster and real-time execution of this simple algorithm requires 
suitable parallelization such as in~\cite{onlinekNN}.

\subsubsection{Perceptron}

The perceptron is an algorithm for learning 
a threshold function: a function that maps its input 
$\mathbf{x} $  (a real-valued vector) to output value 
$f(\mathbf{x}) $ (a single binary value)
  
$$ f(\mathbf {x} )={\begin{cases} 1 & {\text{if }}\ \mathbf {w} \cdot \mathbf {x} +b>0,\\
                                                    
                                                    0&{\text{otherwise}}\end{cases}},
                                                 $$  
where $ \mathbf {w}$  is a vector of real-valued weights, 
$ \mathbf {w} \cdot \mathbf {x}  = \sum _{i=1}^{m}w_{i}x_{i}$
where $m$ is the number of inputs to the perceptron, and $b$ is the bias 
which shifts the decision boundary away from the origin and is 
independent from the input. 

The perceptron learning algorithm needs exponential time, though there are 
algorithms requiring $O(n^{7/2})$ steps. 

A natural extension of the single-layer perceptron described above is the multilayer 
perceptron which essentially contains many perceptrons organized into layers 
thereby gaining in the ability to solve more complex problems in reasonable time.

\subsubsection{Neural networks for real-time object detection }

A reliable and highly accurate real-time object detection algorithm 
is of  paramount importance in defense. 
Its goal is to detect instances of semantic objects of a certain class (such as humans, animals, roads, or vehicles)
in a given video in real-time. 
The input is given in the form of a continuous video stream, and the output is given as a tuple of annotated 
descriptors assigned to each detected object  bounded by an appropriate box in the frame 
in which it appears. 

The YOLO algorithm, presented in~\cite{YOLO} 
is considered state of the art 
algorithm. The processing pipeline of YOLO 
 comprises a single neural network which first predicts bounding boxes 
 in images, after which the problem is reduced to regression on spatially separated 
 bounding boxes. 
 
The military application of this approach imposes two additional constraints: 
(1) no trade-offs between  
accuracy and speed, and (2) not using  networks pretrained outside
 the security 
perimeter for security reasons. 

These constraints and the nature of the problem require huge 
continuous processing power 
over a flow of data for these algorithms to be used.

\subsubsection{Support Vector Machine (SVM) } 

To detect intrusions (IDS) in a network,  its traffic is analyzed for particular 
signatures. Normal network traffic often exhibits a similar signature to 
attacks, and hackers often apply obfuscation for network intrusion. 

Machine learning offers a wide range of efficient tools for 
accurately identifying an IDS, with a restricition that
 training datasets should not 
be associated with malicious data. 
The  
Support Vector Machine (SVM) is a promising candidate for this 
task~\cite{SVMHack}. 

 This algorithm  aims to find a hyperplane in the $N$--dimensional space that 
separates data points while keeping the maximum margin, that is, maximum distance 
between the points of individual classes. 
Formally, suppose that we are given a training dataset of $n$ points 
$(\vec{x}_1, y_1), (\vec{x}_2, y_2), \dots, (\vec{x}_n, y_n)$, where $y_i\in\{-1,1\}$ denotes 
the class that $\vec{x}_i$ belongs to. 
The goal is to find the hyperplane which divides points $\vec{x}_i$ for which $y_i=1$
from those for which $y_i=-1$, whereas the distance from the nearest point $\vec{x}_i$ 
to the hyperplane is maximized. 

Even though the space and the time complexity of SVM are 
polynomial (quadratic and qubic on the size of the input 
respectively), the amount of  data in a network calls for special 
architecture for this algorithm to be efficiently used for the aforementioned 
purpose.

\subsubsection{k-means}

k-means clustering is a method of vector quantization, 
 that aims to partition $n$ observations into $k$ clusters, whereby 
  each observation 
 belongs to the cluster with the nearest mean, serving as a prototype of the cluster. 
 This results in a partitioning of the data space into Voronoi cells. 

Given a set of observations $\textbf{x}=(\vec{x}_1, \vec{x}_2, ..., \vec{x}_n)$, 
where each observation is a $d$-dimensional real vector, 
k-means clustering aims to partition $n$ observations into $k,  k<=n$ 
sets $S = \{S_1, S_2, ..., S_k\}$ 
so as to minimize variance within the cluster. 

Formally, the objective is to find:
$$ \arg\min_{S}{\sum_{i=1}^k \sum_{x\in S_i} \| \textbf{x} - \mu_i\|^2},$$ 
where $\mu_i$ is the mean of points in $S_i$. 

The problem is computationally difficult (NP-hard); however, efficient heuristic algorithms 
quickly converge  to the local optimum for most problem instances.

\subsubsection{Ensemble modeling -- boosting}
Ensemble modeling uses multiple different modeling algorithms or 
different training datasets to predict an outcome. 
The ensemble model then aggregates the prediction of every used model and 
derives its final prediction for unseen data. 
In particular, boosting is a kind of ensemble modeling that has  been widely 
used in military applications (see~\cite{adaBoost}).

A \emph{boost classifier} is a classifier 
$$ F_{T}(x)=\sum _{t=1}^{T}f_{t}(x)\,\!$$
where each $f_{t}$ is a ``weaker'' classifier that takes 
object  $\mathbf {x}$ 
as input and returns a value indicating the class of the object. 

Each of these classifiers produces an output hypothesis, $h(x_i)$, 
for each sample in the training set. 
At each iteration $t$, a weak classifier is selected and assigned  
coefficient
$\alpha_t $ such that the sum of the training error $E_{t}$ 
of the resulting $t$-stage boost classifier is minimized.

In practice, AdaBoost algorithm is realized by cascading the number 
of SVM weak classifiers described 
above.

\section{Dataflow}

The dataflow paradigm~\cite{Amilutinovic2015guide, Cmilutinovic1987open, Dflynn2013moving}  
has been introduced as contrast to the
traditional controlflow paradigm~\cite{Btomasevic1994hardware}. 
In controlflow, a program is
written with the intention to micro-control the flow of data through 
hardware. In dataflow, a program is written with the intention to
configure  hardware, so that, in the ideal case, voltage
difference can move data through hardware.

The dataflow paradigm can achieve speedups of 10x, 100x, or even
1000x, compared to the controlflow paradigm. At the same time, power
reduction can be about 10x. Precision can be varied throughout the
algorithm, which saves chip area. The size of equipment also gets
reduced with a factor of up to 10x.

The algorithms that benefit the most from this paradigm are those
characterized by time-consuming loops and lots of data usability
within each particular loop iteration. Among the algorithms surveyed in
this article, the most suitable ones for dataflow implementation are:
logistic regression, k-means, and ensemble modeling.

Examples of dataflow implementations of these algorithms, as well as
other similar ones, can be found at {\tt appgallery.maxeler.com} or in~\cite{Amilutinovic2015guide}.
For more information,  interested readers are referred to
references~\cite{Btomasevic1994hardware, Cmilutinovic1987open, Dflynn2013moving}.

\section{GaAs}
GaAs technology can also be used for design of processors and
implementation of algorithms. It offers significantly higher processor speed and
level or radiation hardness, which makes it suitable for use in
aerospace and defense environments. On the other hand, the number of
transistors that can be placed on a single chip is smaller, while gate delay heavily 
depends on the gate fan-out.

These characteristics define the specific requirements of processor
design and algorithm implementation. On one hand, not much logic could
be placed onto a single chip, and on the other hand, the ratio of
off-chip to on-chip delays is relatively high. This mandates the
utilization of highly pipelined architectures in which pipeline elements
are of a relatively small complexity.

Efforts to implement various types of processors, under DARPA
sponsorship, were described in~\cite{amilutinovic1986guest} and~\cite{c-helbig1989dcfl}. 
Important concepts  were
described in~\cite{bmilutinovic1987gaas} and~\cite{d-milutinovic1986introduction}. 
These concepts are also relevant also for the
implementation of algorithms described in this article.

Based on the facts given above, it follows that the most effective
implementations can be expected from algorithms that can be
implemented using many small elements, connected in a pipelined fashion like 
image/video registration and stitching. 
Other examples include perceptron, SVM, k-means, and ensemble modeling.
These statements were verified through a number of student projects at the
universities where the coauthors of this article  teach.

 \section{Conclusions}
The surveyed algorithms were chosen by the frequency of usage in selected military 
applications. 
They  have been studied from the viewpoint 
of their implementation based on the dataflow paradigm and GaAs technology.

It has been found that some algorithms are better suited for dataflow than others. 
Namely, the best suited ones are those characterized by a high contribution of loops 
to the overall run time, as well as those with a high level of data reusability within 
each loop iteration.

As far as the potential benefits coming from the utilization of GaAs technology, 
the best performance increases are expected from those algorithms that can be 
implemented on a large number of small modules, connected in a pipelined or systolic fashion. 
Furthermore, the algorithms being less sensitive to large ratios of off-chip to on-chip delays are better suited 
for this technology, which offers high speedups,  but does not permit large chips.

Finally, this survey opens new research avenues related to the synergies in the triangle: 
algorithms -- architectures -- technologies. 
To make an appropriate selection of a specific algorithm from a plethora of options, 
it is neccessary to conduct an analysis of the kind 
presented in this article.

\bibliographystyle{unsrt}

 \bibliography{vmRadBIB}

\end{document}